\documentclass[preprint]{aastex}
\usepackage{geometry}
\usepackage{graphicx}
\usepackage{epsfig}

\newcommand{\Msun}{\ensuremath{\textrm{M}_{\odot}}}

\newcommand{\kms}{\textrm{km\hspace{0.25em}s$^{-1}$}}

\newcommand{\Cofs}{\ensuremath{^{56}}\textrm{Co}}
\newcommand{\Nifs}{\ensuremath{^{56}}\textrm{Ni}}

\begin{document}

\title{Ultraviolet Spectroscopy of Type IIb Supernovae: Diversity and the Impact of Circumstellar Material}
\author{Sagi Ben-Ami\altaffilmark{1,2,3}, Stephan Hachinger\altaffilmark{4,5}, 
Avishay Gal-Yam\altaffilmark{2,6}, Paolo A. Mazzali\altaffilmark{7,4,8},
Alexei V. Filippenko\altaffilmark{9}, Assaf Horesh\altaffilmark{2}, 
Thomas Matheson\altaffilmark{10}, Maryam Modjaz\altaffilmark{11},
Daniel N. Sauer\altaffilmark{12}, Jeffrey M. Silverman\altaffilmark{13},
Nathan Smith\altaffilmark{14}, and Ofer Yaron\altaffilmark{2}}

\altaffiltext{1}{Smithsonian Astrophysical Observatory, Harvard-Smithsonian Center for Astrophysics, 60 Garden St., Cambridge, MA 02138, USA}
\altaffiltext{2}{Department of Particle Physics and Astrophysics, The Weizmann Institute of Science, Rehovot 76100, Israel}
\altaffiltext{3}{email: sbenami@cfa.harvard.edu}
\altaffiltext{4}{Max-Planck-Institut f\"ur Astrophysik, Karl-Schwarzschild-Str. 1, 85748 Garching, Germany}
\altaffiltext{5}{Institut f\"ur Theoretische Physik und Astrophysik, Universit\"at Wurzburg, Emil-Fischer-Str. 31, 97074 Wurzburg, Germany}
\altaffiltext{6}{Kimmel Investigator}
\altaffiltext{7}{Astrophysics Research Institute, Liverpool John Moores University, Liverpool L3 5RF, United Kingdom}
\altaffiltext{8}{INAF-Osservatorio Astronomico, vicolo dell'Osservatorio, 5, I-35122 Padova, Italy}
\altaffiltext{9}{Department of Astronomy, University of California, Berkeley, CA 94720-3411, USA}
\altaffiltext{10}{National Optical Astronomy Observatory, 950 N. Cherry Avenue, Tucson, AZ 85719, USA}
\altaffiltext{11}{Center for Cosmology and Particle Physics, Department of Physics, New York University, 4 Washington Place, Room 529, New York, NY 10003, USA}
\altaffiltext{12}{Department of Astronomy, Stockholm University, Albanova University Center, SE-106 91 Stockholm, Sweden}
\altaffiltext{13}{Department of Astronomy, University of Texas, Austin, TX 78712, USA}
\altaffiltext{14}{Steward Observatory, University of Arizona, Tucson, AZ 85721, USA}

\begin{abstract}
We present new {\it Hubble Space Telescope} ({\it HST}) multi-epoch ultraviolet (UV) spectra of 
the bright Type IIb SN~2013df, and undertake a comprehensive analysis of the set of 
four Type IIb supernovae for which {\it HST} UV spectra are available (SN~1993J, SN~2001ig, SN~2011dh, 
and SN~2013df). We find strong diversity in both continuum levels and line features among these
objects. We use radiative-transfer models that fit the optical part of the spectrum well, and find that
in three of these four events we see a UV continuum flux excess, apparently unaffected by line absorption. 
We hypothesize that this emission originates above the photosphere, and is related
to interaction with circumstellar material (CSM) located in close proximity to the SN progenitor.   
In contrast, the spectra of SN~2001ig are well fit by single-temperature models, display weak continuum and strong 
reverse-fluorescence features, and are similar to spectra of radioactive $^{56}$Ni-dominated Type Ia supernovae. 
A comparison of the early shock-cooling components in the observed light curves
with the UV continuum levels which we assume trace the strength of CSM interaction suggests that
events with slower cooling have stronger CSM emission. 
The radio emission from events having a prominent UV excess is perhaps consistent with 
slower blast-wave velocities, as expected if the explosion shock was slowed down by the CSM that
is also responsible for the strong UV, but this connection is currently speculative as it is based
on only a few events. \end{abstract}

\vspace{0.5cm}
\section{Introduction}

A  core-collapse supernova (CC~SN) occurs when a massive star ($\gtrsim 8$\,M$_{\odot}$) ends its life in a terminal explosion that destroys the star completely and leaves a neutron star or a stellar-mass black hole as a compact remnant (e.g., Woosley \& Janka 2005). Classification of such supernovae (SNe) is based primarily on their observed optical spectra, but also on their observed light curves (LCs; e.g., Type IIP vs. Type IIL SNe). Type II SNe have obvious hydrogen Balmer lines throughout their evolution, whereas SNe~Ib lack prominent signatures of H while having strong He lines, and SNe~Ic lack clear signatures of both H and He (e.g., Filippenko 1997). Type IIb SNe populate a unique location between the H-rich Type II and the stripped-envelope Type Ib/Ic SNe (Filippenko 1988, 1997). 
 
Early-time spectra of Type IIb events are similar to those of SNe~II, with strong Balmer lines dominating the optical spectrum. Later (i.e., days to weeks; Filippenko et al. 1993, Crockett et al. 2008; Bufano et al. 2014), the spectra evolve to resemble those of SNe~Ib, with prominent He~I lines. 
This behavior is attributed to the presence of a low-mass H-rich envelope --- e.g., 0.18\,M$_{\odot}$ for SN~1993J, the best-studied SN~IIb to date (Woosley et al. 1994; Hachinger et al. 2012).  
Two common mechanisms that are invoked to explain the partial stripping of SN~IIb progenitors are stellar winds from a massive star (e.g., SN~2001ig; Ryder et al. 2004) and mass loss through Roche-lobe overflow in a binary system (e.g., SN~1993J; Maund et al. 2004).      
The LCs of SNe~IIb resemble those of radioactively powered SNe~Ib \cite{Arcavi2012}. In some cases, two peaks are observed in both the optical and the ultraviolet (UV; e.g., Filippenko et al. 1993; Arcavi et al. 2011; Van Dyk et al. 2014). The first peak is attributed to the shock-cooling phase, and the second to radioactive nickel decay (e.g., Nakar \& Piro 2014). 

For some nearby SNe~IIb, analysis of {\it Hubble Space Telescope (HST)} images taken prior to the SN explosion led to the identification of the progenitor stars. 
In a few cases, namely SN~1993J, SN~2011dh, and SN~2013df, the progenitors were yellow supergiants (YSGs; $R=150$--650\,R$_{\odot}$; Maund et al. 2004, 2011; Van Dyk et al. 2011, 2014), somewhat hotter ($T \approx 6000$\,K) than red supergiants, the progenitors of the more common Type IIP SNe \cite{Smartt2009}. For SN~1993J, Maund et al. (2004) and Fox et al. (2014) find evidence for the presence of a hot (B2\,Ia) star --- a possible companion of the progenitor. Georgy (2012) shows, using stellar models including rotation, that 12--15\,M$_{\odot}$ main-sequence stars can end their lives as YSGs following increased mass-loss rates during their red supergiant phase. Smith et al. (2011) find that the statistics of SN subtypes require that binary interaction is a primary channel for stripped-envelope SNe. 

Stancliffe \& Eldridge (2009) model the proposed binary system of the SN~1993J progenitor, in an attempt to understand whether stripping by a companion in a binary system might result in one of the stars becoming a YSG. In the case of SN~2011dh, Bersten et al. (2012) claim that evolution in a binary system has most likely caused the stripping as well.
Using the ``flash spectroscopy'' method, Gal-Yam et al. (2014) demonstrate that the progenitor of the Type IIb SN 2013cu (iPTF13ast) likely had a surface composition similar to that of a late-type WN Wolf-Rayet (W-R) star (Smith \& Conti 2008).

Outside of the optical waveband, studies in the radio and X-ray bands have probed the mass-loss history of these events in an attempt to uncover the progenitor stripping mechanism (e.g., Van Dyk et al. 1994; Fransson et al. 1996; Ryder et al. 2004; Chevalier \& Fransson 2006). Indications of circumstellar material (CSM) with a density flatter than $\rho_w \propto r^{-2}$ have been provided by radio-light-curve modeling, and X-ray emission from shocked CSM was found to be responsible for most of the X-ray flux at early times \cite{Fransson1996}. Ryder et al. (2004, but see Soderberg et al. 2006) postulate that the observed modulations in the radio LC of SN~2001ig are produced by density modulations in the CSM caused by motion in an eccentric binary system.

Chevalier \& Soderberg (2010) suggest that SNe~IIb can be further divided into two subgroups. The first includes explosions of compact objects (Type cIIb, $R \approx 1$\,R$_\odot$). These events are distinguished by weak optical emission from the shock-heated envelope at early times (i.e., adiabatic cooling phase), rapid expansion of the radio shell because of low CSM density resulting from high wind velocities, and nonthermal X-ray emission from inverse-Compton scattering \cite{Fransson1996}. SN~2001ig and SN~2008ax are postulated to be members of this subgroup. In the case of SN~2008ax, the classification is also supported by the identification of a possible compact W-R progenitor in pre-explosion {\it HST} images \cite{Crockett2008}, even though this conclusion is not as strong as in the cases of SN~1993J, SN~2011dh, and SN~2013df  (a higher confidence level in the detection of the progenitor in pre-explosion images taken with \textit{HST}). The second subgroup originates from explosions of cool extended progenitors surrounded by dense winds (Type eIIb, $R \gtrsim 100$\,R$_\odot$). SN~1993J is the best-studied member of this group. 

Despite the large number of investigations focused on SNe~IIb, not much attention has been paid to the UV emission of these events, despite the flux excess some of them have exhibited compared to predicted theoretical models (e.g., Baron et al. 1993; Jeffery et al. 1994). This is partially because of the small number of events for which we have high-quality datasets in the UV band. In the following work we present early-time UV spectroscopy taken with {\it HST} for a sample of SNe~IIb, with a focus on the recent SN~2013df, finding that UV spectra further illuminate the diversity observed in SNe~IIb. We argue that the UV spectral energy distribution (SED) is governed by two parameters: the amount of CSM in the object's vicinity at the time of explosion, and the amount of $^{56}$Ni synthesized and ejected in the explosion. 
We suggest that future division of Type IIb SNe will benefit by taking into account the observed UV SED as an indicator of the presence of CSM around the progenitor, introducing another level of complexity to the picture suggested by Chevalier \& Soderberg (2010). 

Section 2 describes the SN sample and observations used in this work, and the
bolometric LCs are assembled. In \S3 we analyze the bolometric LCs, identify dominant features in the observed spectra, construct radiative-transfer models, produce synthetic spectra, and compare these to {\it HST} observations. We present our conclusions and discuss their implications in \S4.

\newpage

\section{Observations}
\subsection{The Sample}
We analyze four SNe~IIb having high signal-to-noise ratio (S/N) UV spectra obtained with {\it HST} within 30\,days after the estimated explosion. This is an exhaustive list; to the best of our knowledge, no other early-time UV spectra of SNe~IIb have been obtained with {\it HST}. 

SN~2013df was discovered on 7.78 June 2013 (Ciabattari et al. 2013; UTC dates are used throughout this paper) in an outer spiral arm of NGC 4414 (distance $d = 16.6\pm0.4$\,Mpc; Freedman et al. 2001). A comparison of $V$-band light curves of SN~1993J, SN~2011dh, and SN~2013df led Van Dyk et al. (2014) to estimate an explosion date of 4 June 2013, which we adopt throughput this paper (see also Morales-Garoffolo et al. 2014). Van Dyk et al. (2014) identify the SN progenitor in pre-explosion {\it HST} images, arguing for an extended star with an effective radius of $545 \pm 65$\,R$_{\odot}$ and an effective temperature of $4250 \pm 100\,$K. The progenitor is likely a YSG with an initial mass of 13--17\,M$_{\odot}$ .

SN~1993J was discovered by amateur astronomer F. Garcia on 28.9 March 1993, at $d = 3.6$\,Mpc in the spiral galaxy M81 (Freedman et al. 1994). We adopt an explosion date of 27.5 March 1993 \cite{Filippenko1993}. The SN~1993J progenitor was identified in ground-based pre-explosion images by Aldering et al. (1994). The progenitor is a YSG with a radius of $\sim 600$\,R$_{\odot}$ and a mass of 12--22\,M$_{\odot}$ \cite{Aldering1994,Maund2004,Fox2014}.
Maund et al. (2004) use photometric and spectroscopic observations taken 10\,yr after the explosion to give evidence for a massive star at the SN location, claimed to be the binary companion of the progenitor. 

SN~2001ig was discovered on 10.43 December 2001 by Evans et al. (2001) in the spiral galaxy NGC 7424 at $d = 11.5$\,Mpc \cite{Soria2006}. 
We adopt an explosion date of 3 December 2001 based on radio LC modeling \cite{Ryder2004}. Using the same modeling, Ryder et al. (2004) argue that the most likely progenitor of SN~2001ig is a W-R star with a radius of $\sim 10^{11}\,$cm. SN~2001ig is the only SN~cIIb candidate for which early-time UV spectra were obtained with {\it HST}.

SN~2011dh was discovered by amateur astronomer A. Riou on 31.83 May 2011 (Arcavi et al. 2011) in the Whirlpool galaxy (M51) at $d = 8.05 \pm 0.35$\,Mpc (Marion et al. 2014). SN~2011dh was not detected down to $g=21.44$\,mag in Palomar Transient Factory images taken on 31.27 May 2011, and we adopt an explosion date of 31.5 May 2011 \cite{Arcavi2011,Marion2014}. Bersten et al. (2012) and Horesh et al. (2013) argue for an extended progenitor with a radius of $\sim 200$\,R$_{\odot}$. This was verified by Van Dyk et al. (2013), who determined that the YSG at the SN location vanished in post-explosion {\it HST} images; see also Maund et al. (2011) and Ergon et al. (2014a,b).

We compare our LCs with those of two additional objects, SN~2008ax \cite{Pastorello2008, Crockett2008} and SN~2011hs \cite{Bufano2014}, albeit these events have no early-time UV spectroscopy.

SN~2008ax was discovered independently by Mostardi, Li, and Filippenko on 3.95 March 2008, and by Nakano and Itagaki on 5.12 March 2008 \cite{Nakano2008}, in the barred spiral galaxy NGC 4490 at $d = 11.1$\,Mpc. Extremely tight limits on the explosion time of this event are placed by Arbour (2008), who monitored the host galaxy only 6\,hr before detection. We adopt an explosion date of 3.7 March 2008 (Pastorello et al. 2008, and references therein). SN~2008ax is a candidate Type cIIb event \cite{Chevalier2010}.

SN~2011hs was discovered on 12.5 November 2011 in the galaxy IC~5267 at $d = 26.4$\,Mpc. We adopt an explosion date of 6.5 November 2011 based on a comparison of early-time photometric data with the SN~1993J LC (Bufano et al. 2014, and reference therein). Bufano et al. (2014) argue for an extended progenitor similar to the one observed for SN~1993J; see \S3.1.

Table 1 lists all of these objects.

\begin{deluxetable}{ccccccc}
\tabletypesize{\tiny}
\tablecaption{Correlations Between UV SED and Other Observables in SNe~IIb}
\tablenum{1}
\tablehead{\colhead{Object Name} & \colhead{Probable Progenitor} & \colhead{UV SED} & \colhead{UV Flux} & \colhead{Shock Cooling}  & \colhead{Blast-Wave} & \colhead{$^{56}$Ni Mass} \\
\colhead{} & \colhead{} & \colhead{Shape} & \colhead{Excess} & \colhead{Phase Duration}  & \colhead{Velocity$^a$} & \colhead{(ejected)} } 
\startdata
SN~1993J   &  YSG ($R\approx600$\,R$_{\odot}$)$^b$     &  quasi-continuum  &  High    &   $\sim\,10$\,days  &   $\sim$\,10,000\,km\,s$^{-1}$   &$<0.1$\,M$_{\odot}$  \\
SN~2001ig  &  Compact ($R\approx1$\,R$_{\odot}$)$^c$  &  line-dominated     &  None   &    not available     &   $\sim$\,50,000\,km\,s$^{-1}$  &$>0.1$\,M$_{\odot}$ \\ 
SN~2011dh &  YSG ($R\approx200$\,R$_{\odot}$)$^d$     &  quasi-continuum  &  Small   &   $\sim$\,3\,days    &   $\sim$\,50,000\,km\,s$^{-1}$  &$<0.1$\,M$_{\odot}$ \\
SN~2013df  &  YSG ($R\approx600$\,R$_{\odot}$)$^e$     &  quasi-continuum  &  High    &    $\sim$\,10\,days  &   $\sim$\,10,000\,km\,s$^{-1}$  &$<0.1$\,M$_{\odot}$ \\ \hline
SN~2008ax  &  Compact ($R\approx1$\,R$_{\odot}$)$^f$     &  ? &  ?    &    $\leq$\,6\,hr  &   $\sim$\,70,000\,km\,s$^{-1}$  &$\sim$\,0.06\,M$_{\odot}$ \\
SN~2011hs  &  YSG ($R\approx600$\,R$_{\odot}$)$^g$   &  ? &  ?    &    $\sim$\,8\,days  &   $\sim$\,50,000\,km\,s$^{-1}$  &$<0.09$\,M$_{\odot}$ \\
                     
\enddata
\tablecomments{See text for references. No early-time photometric data exist for SN~2001ig. However, SN~2008ax, which is also a SN~cIIb candidate, shows no shock-cooling phase as early as 6\,hr after the estimated explosion (Crockett et al. 2008).\\
$^a$e.g., Crockett \& Soderberg (2010), Horesh et al. (2013), Bufano et al. (2014).\\
$^b$Aldering et al. (1994), Van Dyk et al. (2002).\\
$^c$Ryder et al. (2004, 2006).\\
$^d$Bersten et al. (2012), Horesh et al. (2013), Van Dyk et al. (2013).\\
$^e$Van Dyk et al. (2014).\\
$^f$Crockett et al. (2008), Crockett \& Soderberg (2010).\\
$^g$Bufano et al. (2014).
}
\end{deluxetable}

\newpage
\subsection{{\it HST} Spectra}
We observed SN~2013df with {\it HST} on four epochs: 17.2, 22.1, 26.1, and 30.2 June 2013 (respectively $\sim 13$, 18, 22, and 26\,days after the estimated explosion date; Program GO-13030, PI A.~V. Filippenko). At each visit, 3 spectra of the SN were obtained: a UV spectrum using STIS/MAMA centered at 2376\,\AA, and two spectra using STIS/CCD centered at 4300\,\AA\ and 7751\,\AA. Data were reduced using the \textit{calstis} software \cite{Bostroem2011}.

A UV+optical spectrum of SN~1993J was obtained on 15.4 April 1993 using {\it HST}-FOS ($\sim19$\,days after the estimated explosion date; Program GO-4528, PI R. P. Kirshner; Jeffery et al. 1994). Data were reduced using the \textit{calfos} software \cite{Keyes1995}.

UV+optical spectra of SN~2001ig were taken on two different epochs, 14 and 22.6 December 2001, using {\it HST}-STIS ($\sim11$ and $\sim19$\,days after the estimated explosion date; Program GO-9114, PI R. P. Kirshner; Marion et al. 2014). Data were reduced using the \textit{calstis} software \cite{Bostroem2011}. 

A UV+optical spectrum of SN~2011dh was obtained with {\it HST}-STIS/CCD on 24.1 June 2011 ($\sim24$\,days after the estimated explosion date; Program GO-12540, PI R. P. Kirshner; Marion et al. 2014).  Data were reduced using the \textit{calstis} software \cite{Bostroem2011}.

Table 2 summarizes all {\it HST} spectroscopic observations used in our analysis. 
Figure 1 shows the spectra described above, with insets highlighting the UV.
Digital versions of these spectra are available from WISeREP (Yaron \& Gal-Yam 2012). 
 
\begin{figure}[h!]
\centerline{
\begin{tabular}{cc}
\scalebox{0.32}{\includegraphics{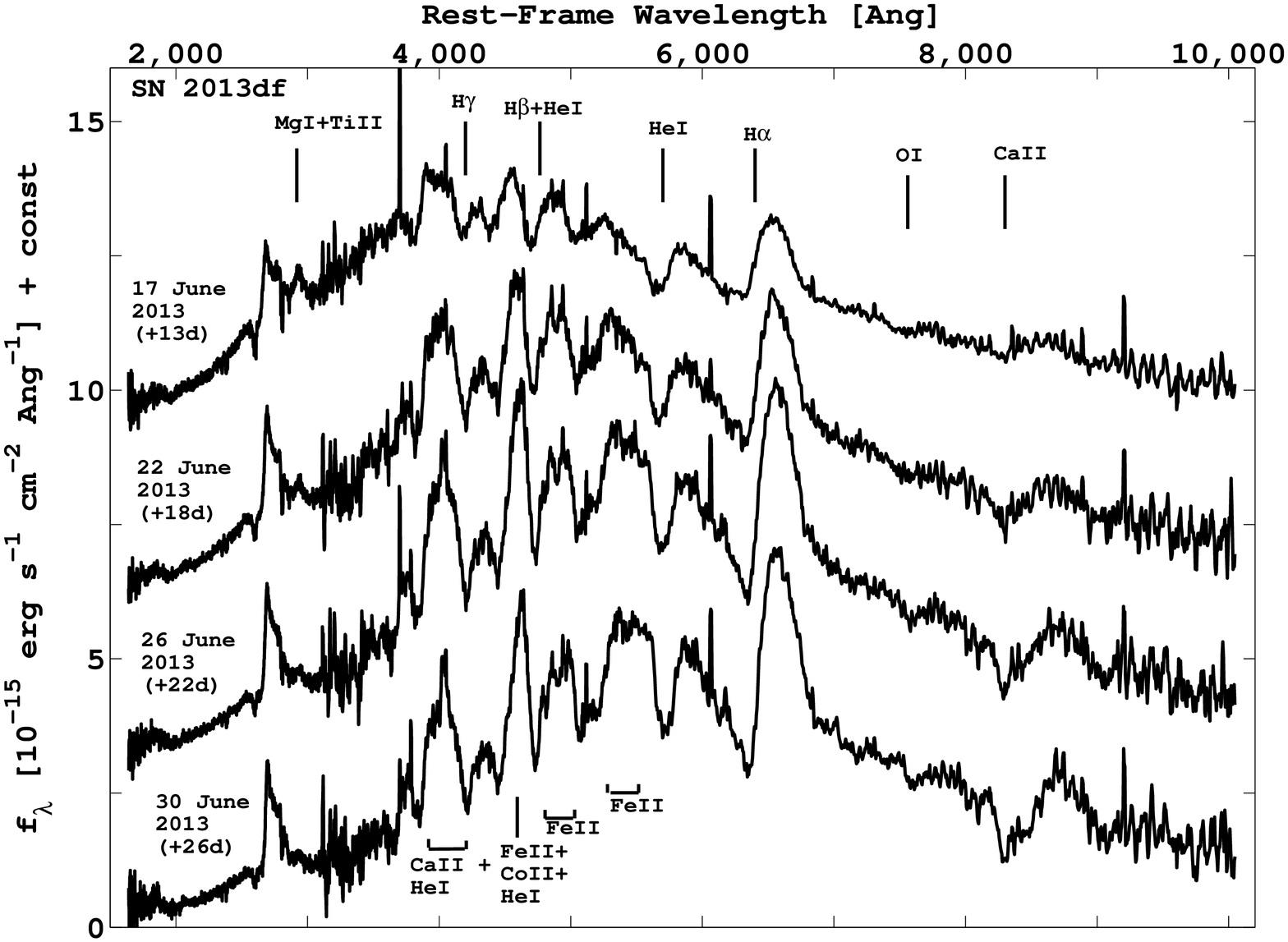}} &
\scalebox{0.32}{\includegraphics{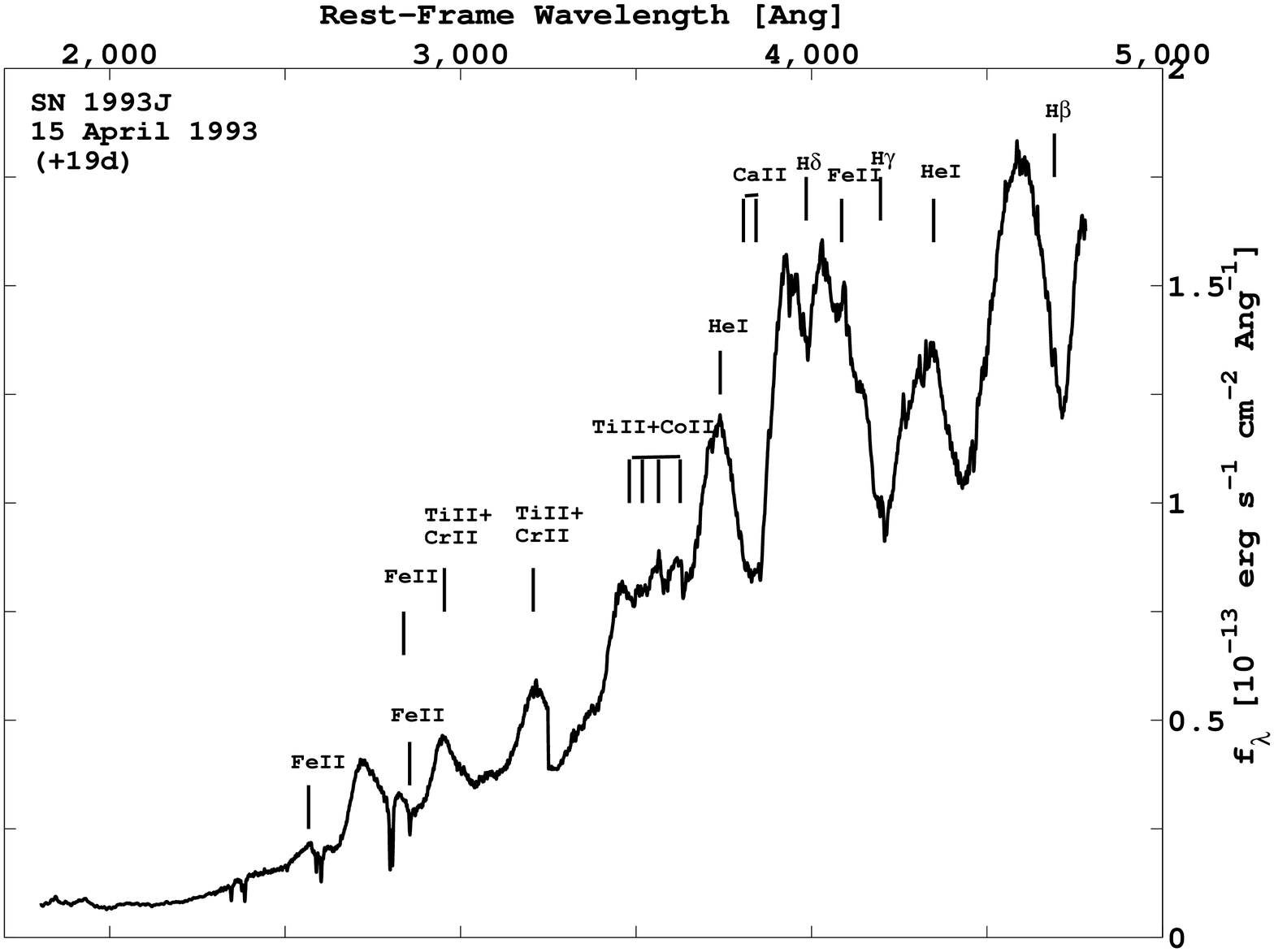}}\\
\scalebox{0.32}{\includegraphics{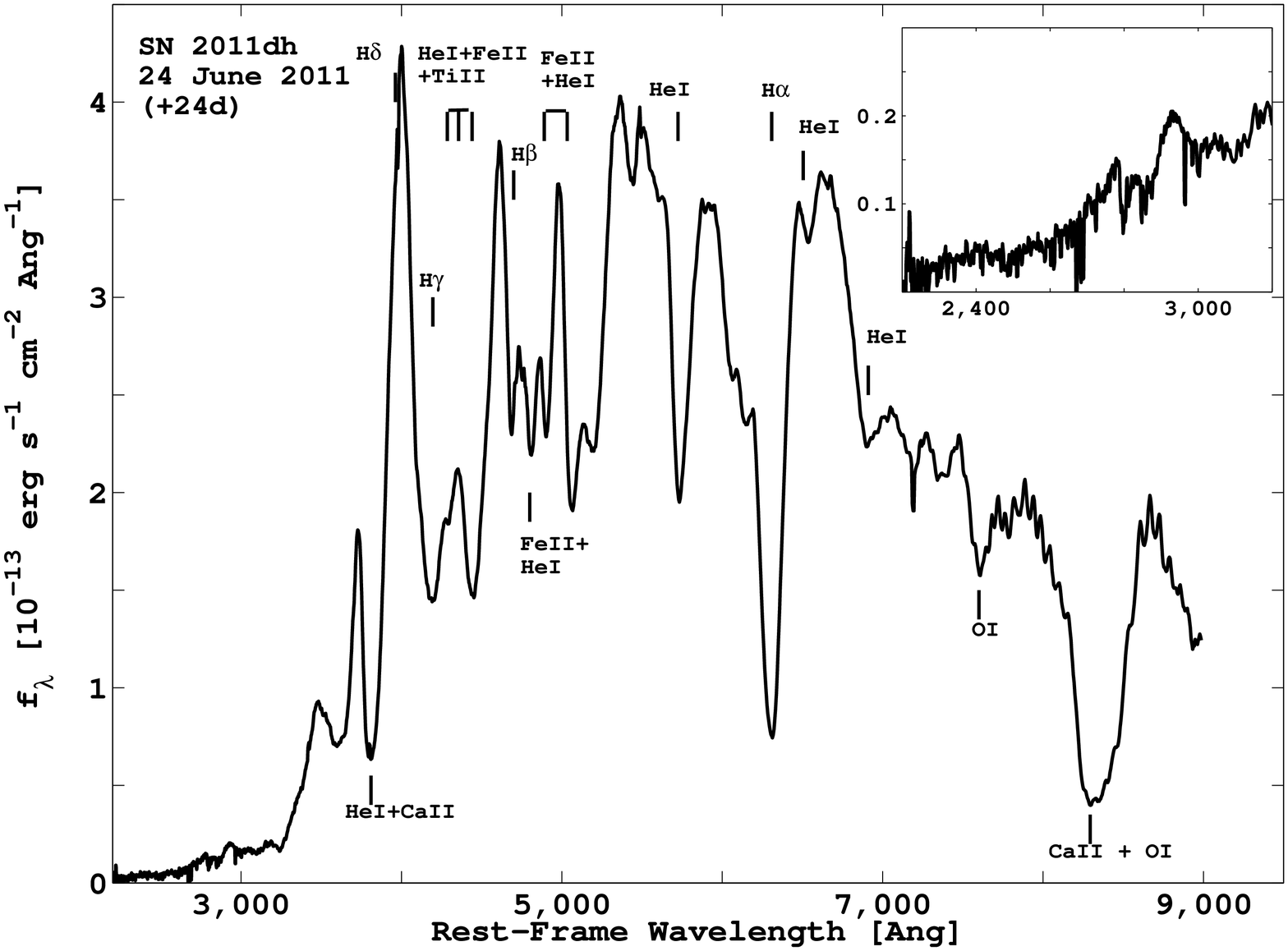}}  &
\scalebox{0.32}{\includegraphics{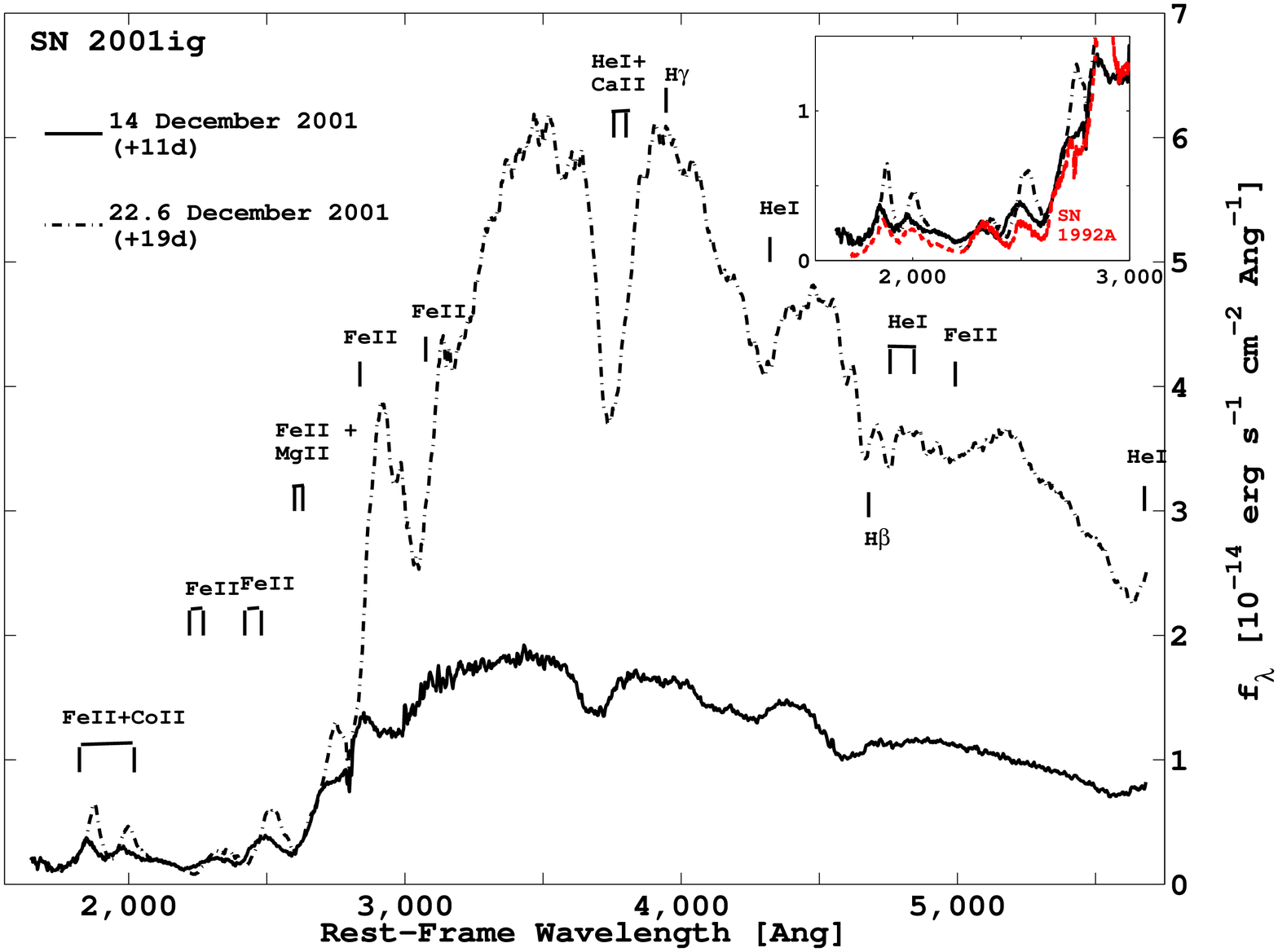}} \\
\end{tabular}}
\caption{\small {\it HST} spectra of our SN~IIb sample. \textit{Top left:} SN~2013df (Program GO-13030, PI A.~V. Filippenko). \textit{Top right:} SN~1993J (Program GO-4528, PI R. P. Kirshner; Jeffery et al. 1994). \textit{Bottom left:} SN~2011dh  (Program GO-12540, PI R. P. Kirshner; Marion et al. 2014). \textit{Bottom right:} SN~2001ig (Program GO-9114, PI R. P. Kirshner; Marion et al. 2014). The inset shows a comparison with the UV spectrum of the Type Ia SN~1992A (Program GO-4016, PI R. P. Kirshner; Kirshner et al. 1993). 
Line identification is based on the spectral model described in \S3.3. In the case of SN~2013df, we can model the strong, broad, asymmetric line at around 2800\,\AA {} using SYNOW (Mg~II, Co~II, Fe~II, and Ti~II at a temperature of 10,000\,K) \citep{Branch2007,Parrent2010}. However, this exercise introduces other lines into the modelled spectra below 2800\,\AA\ which we do not observe. 
Digital data are available from the Weizmann Interactive Supernova Data Repository (WISeREP; Yaron \& Gal-Yam 2012; http://www.weizmann.ac.il/astrophysics/wiserep/).}
\label{fig:HSTSpectra}
\end{figure}  

\begin{deluxetable}{cccccc}
\tabletypesize{\tiny}
\tablecaption{{\it HST} Observations}
\tablenum{2}
\tablehead{\colhead{Object Name} & \colhead{Date} & \colhead{Exposure Time} & \colhead{Instrument} & \colhead{Detector} & \colhead{Aperture/Grating} \\ 
\colhead{} & \colhead{(UT)} & \colhead{(s)} & \colhead{} & \colhead{} & \colhead{} } 
\startdata
  SN~2013df  & 17.2 June 2013  &  8476  &  STIS  &  NUV-MAMA  &  52X0.2/G230L  \\
                     &  17.4 June 2013   &  222  &  STIS  &  CCD &  52X0.2/G430L  \\ 
                     &  17.4 June 2013   &  222  &  STIS  &  CCD &  52X0.2/G750L  \\ 
                     &  22.1 June 2013   &  8594  &  STIS  &  NUV-MAMA  &  52X0.2/G230L  \\
                     &  22.3 June 2013   &  108  &  STIS  &  CCD &  52X0.2/G430L  \\ 
                     &  22.3 June 2013  &  108  &  STIS  &  CCD &  52X0.2/G750L  \\ 
                     &  26.1 June 2013  &  8594  &  STIS  &  NUV-MAMA  &  52X0.2/G230L  \\
                     &  26.3 June 2013  &  108  &  STIS  &  CCD &  52X0.2/G430L  \\ 
                     &  26.3 June 2013  &  108  &  STIS  &  CCD &  52X0.2/G750L  \\ 
                     &  30.2 June 2013  &  6118  &  STIS  &  NUV-MAMA  &  52X0.2/G230L  \\
                     &  30.3 June 2013  &  118  &  STIS  &  CCD &  52X0.2/G430L  \\ 
                     &  30.3 June 2013  &  118  &  STIS  &  CCD &  52X0.2/G750L  \\ 
                     \hline \\
  SN~2011dh  &  24.1 June 2011  &  3600  &  STIS  &  CCD  &  52X0.2/G230LB  \\  
                       &  24.2 June 2011  &  800  &  STIS  &  CCD  &  52X0.2/G430L  \\ 
                       &  24.2 June 2011 &  350  &  STIS  &  CCD  &  52X0.2/G750L  \\ 
                      \hline \\
  SN~2001ig  & 14.0 December 2001 &  4343  &  STIS  &  NUV-MAMA  &  0.2X0.2/E140M  \\  
		    &  14.1 December 2001 &  2703  &  STIS  &  NUV-MAMA  &  52X0.2/G230L  \\
		    &  14.2 December 2001 &  200    &  STIS  &  CCD              &  52X0.2/G430L  \\    
  		    &  22.6 December 2001 &  7907  &  STIS  &  NUV-MAMA  &  0.2X0.2/E140M  \\  
		    &  22.9 December 2001 &   5514  &  STIS  &  NUV-MAMA  &  52X0.2/G230L  \\
		    &  22.6 December 2001 &  300    &  STIS  &  CCD  &  52X0.2/G430L  \\ 
		       
                        \hline \\
  SN~1993J  & 15.4 April 1993  &  1800  &  FOS &   &  4.3/G160L  \\                  
                    & 15.5 April 1993   &  1800  &  FOS  &    &  4.3/G270H  \\
                    & 15.5 April 1993   &  1300  &  FOS  &    &  4.3/G400H  \\  
\enddata
\tablecomments{Digital data are available from the Weizmann Interactive Supernova Data Repository (WISeREP; Yaron \& Gal-Yam 2012; http://www.weizmann.ac.il/astrophysics/wiserep/).\\
}
\end{deluxetable}

\subsection{Bolometric Light Curves}
In order to set the context for the analysis of the UV emission, we first inspect the bolometric LCs of SNe~IIb.
We collected the LC of SN~1993J from Richmond et al. (1996), the LC of SN~2008ax from Pastorello et al. (2008), and the LC of SN~2011hs from Bufano et al. (2014).

\begin{figure}[h!p!]
\center
    \scalebox{0.5}{\includegraphics{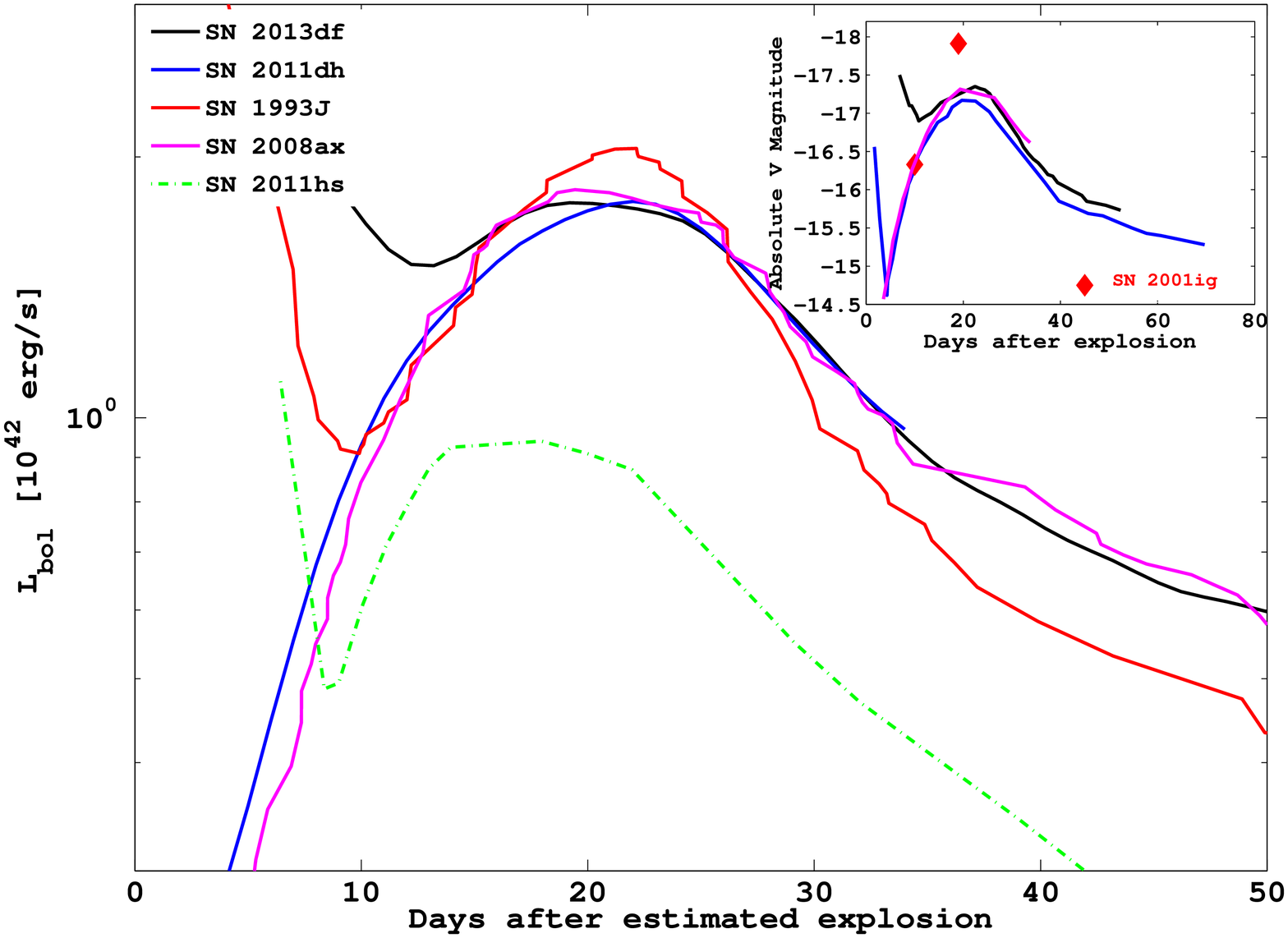}}
  \caption{\small Bolometric LC comparison. 
  Similar behavior is exhibited in all cases. The variations are caused by a different mass of $^{56}$Ni ejected in each event, the shock-breakout radius, and the progenitor density profile \cite{Perets2010,Nakar2014}. Bolometric LCs are taken from Richmond et al. (1996; SN~1993J), Pastorello et al. (2008; SN~2008ax), Marion et al. (2014; SN~2011dh), and Bufano et al. (2014; SN~2011hs). Inset: A $V$-band comparison suggests that SN~2001ig is the intrinsically brightest event in our sample (no bolometric data exist).  The inset also shows the initial shock-cooling phase for SN~2011dh (based on unfiltered measurements by amateur astronomers; see Arcavi et al. 2011).
          }
          \label{fig:Bol_LC}
\end{figure}

In the case of SN~2011dh, the bolometric LC is taken from Marion et al. (2014). This LC starts 4 days after estimated explosion and does not include the initial adiabatic cooling phase. However, Arcavi et al. (2011) show that such a phase was detected using mostly unfiltered photometry from amateur astronomers (see Fig. \ref{fig:Bol_LC} inset). The duration of the shock-cooling phase for SN~2011dh is less than 3\,days \cite{Arcavi2011}; see also Ergon et al. (2014a,b).

We construct a bolometric LC for SN~2013df from observations\footnote{For LCs at specific bands, see Van Dyk et al. (2014).} obtained with {\it Swift}-UVOT (UVW2, UVM2, and UVW1 bands), KAIT ($B$ and $V$ bands), and RATIR ($r$, $i$, $z$, $J$, and $H$ bands). 
Our LC samples the period between 13 June 2013 and 25 July 2013. 
Magnitudes were corrected for host-galaxy and Milky Way extinction assuming $A_V=0.30$\,mag and using the Cardelli et al. (1989) reddening law (see Van Dyk et al. 2014 for a thorough discussion of extinction estimates for SN~2013df). 
Magnitudes were converted to quasi-monochromatic flux at the effective wavelength of each filter. 
A quasi-synthetic light curve at each band was generated by fitting a spline to the measurements. In cases where the observations started on 14 June 2013 (KAIT $B$ and $V$ bands) and June 16 2013 (RATIR $zHK$), an artificial data point was added on 13 June 2013 using backward extrapolation of the initial three measurements in each band. Varying this artificial point value by 20\%, we find that this method introduces an error of $<$\,10\% to the bolometric flux at this time. 
The monochromatic LCs were than summed using the trapezoidal integration method, with bins equal to the full width at half-maximum intensity (FWHM) of each filter to derive a bolometric LC. In view of our well-sampled LC, and based on blackbody fits to the observed {\it HST} spectra, we estimate a coverage of $\geq95$\% of the total bolometric flux, and no attempt was made to correct for emission below 2100\,\AA\ or above 17,800\,\AA. The bolometric LCs are plotted in Figure \ref{fig:Bol_LC}.

\section{Analysis}
\subsection{Light-Curve Comparison}
Comparing the bolometric LCs, we find that, with the exception of SN~2011hs, a similar width around peak magnitude is observed; see Figure \ref{fig:Bol_LC}.
As noticed by Pastorello et al. (2008) for SN~2008ax and SN~1993J, this suggests a similar value for the $M_{\rm ej}^3$/$E_{\rm k}$ ratio according to Arnett's law \cite{Arnett1982}.
The difference in luminosity is attributed to small changes in the amount of $^{56}$Ni ejected in the explosion (e.g., Roming et al. 2009; Perets et al. 2010; Maurer et al. 2010; Ergon et al. 2014a).

The most significant difference between the observed LCs is the early contribution from the shock-heated envelope. Of the four events for which the shock-cooling phase duration can be estimated, SN~2011dh exhibits the fastest decay, followed by SN~2011hs,  SN~1993J, and SN~2013df, in agreement with Morales-Garoffolo et al. (2014).

Nakar \& Piro (2014) argue that for progenitors with a loosely bound envelope and an envelope mass much smaller than the core mass, the cooling-phase duration is proportional to the amount of mass concentrated at the stellar radius. This suggests that the envelope of SN~2013df is more massive than that of, for example, SN~1993J. However, this does not agree with our findings from spectral modeling (Fig. 4), as well as with the conclusions of Morales-Garoffolo et al. (2014). We therefore suggest that perhaps an extended shock-cooling phase might also be caused by the presence of circumstellar material at the vicinity of the progenitor; see \S3.2 and \S4. This scenario is also supported by an analysis of nebular spectra presented by Morales-Garoffolo et al. (2014).
In the case of SN~2008ax, the absence of a shock-cooling phase, despite early photometric measurements taken $<$\,6\,hr from the time of explosion, led Pastorello et al. (2008) and Chevalier \& Soderberg (2010) to conclude  that it had a compact progenitor ($R\approx1$\,R$_{\odot}$). We note that this also likely means it did not have dense CSM, and may also argue against a nearby binary companion \cite{Crockett2008}.  

Two events that sample the lower and upper energy regimes in our sample are SN~2011hs and SN~2001ig \cite{Ryder2004,Ryder2006,Bufano2014}.
In the case of SN~2011hs, the estimated $^{56}$Ni mass and the energy are 0.04\,M$_{\odot}$ and $8\times10^{50}$\,erg, respectively. An extended progenitor ($R\approx550$\,R$_{\odot}$) is most likely based on the observed initial decline in the LC \cite{Bufano2014}.
For SN~2001ig, no reliable photometric data were found. We therefore integrate over the {\it HST} spectra in the wavelength range 5000--5800\,\AA\ to create two data points in the $V$ band, and compare this result to the $V$-band LC of SN~2008ax and SN~2013df (Fig. \ref{fig:Bol_LC}, inset). SN~2001ig is the most luminous event in our sample in $V$. Silverman et al. (2009) suggest that 0.13\,M$_{\odot}$ of $^{56}$Ni was ejected by this event, slightly lower than predicted by the peak $V$ vs. Ni-mass correlation of Perets et al. (2010), if we assume the second data point represents the SN peak magnitude. The effect of the higher luminosity on the UV emission from this event is discussed in \S3.2.

\subsection{Observed Spectra}
SN~2013df spectra longward of $\sim 4000$\,\AA\ exhibit a continuum with strong absorption lines, as expected in the photospheric phase of CC~SNe.
The spectra are dominated\footnote{Line identification is based on the spectral model described in \S3.3.} by H Balmer lines (6563, 4861, 4340\,\AA), He~I lines (5875, 4471, 3888\,\AA), Ca~II H\&K (3968, 3934\,\AA), the Ca~II near-infrared triplet (8662, 8542, 8498\,\AA), the O~I triplet (7774\,\AA), and many features from iron-group elements (e.g., Fe~II lines at 5316, 5276, 4583\,\AA\ and Co~II at 4660\,\AA). Below $\sim 4000$\,\AA, all spectra show an excess of emission with respect to a single-temperature model. The SED at these wavelengths is smooth, with a strong, broad, asymmetric line at around 2800\,\AA. Using SYNOW \citep{Branch2007,Parrent2010}, we can model the observed line profile with Mg~II, Co~II, Fe~II, and Ti~II at a temperature of 10,000\,K. However, this exercise introduces other lines into the modelled spectra below 2800\,\AA\ which we do not observe. 

\begin{figure}[h!p!]
\scalebox{0.4}{\includegraphics[angle=270,origin=c]{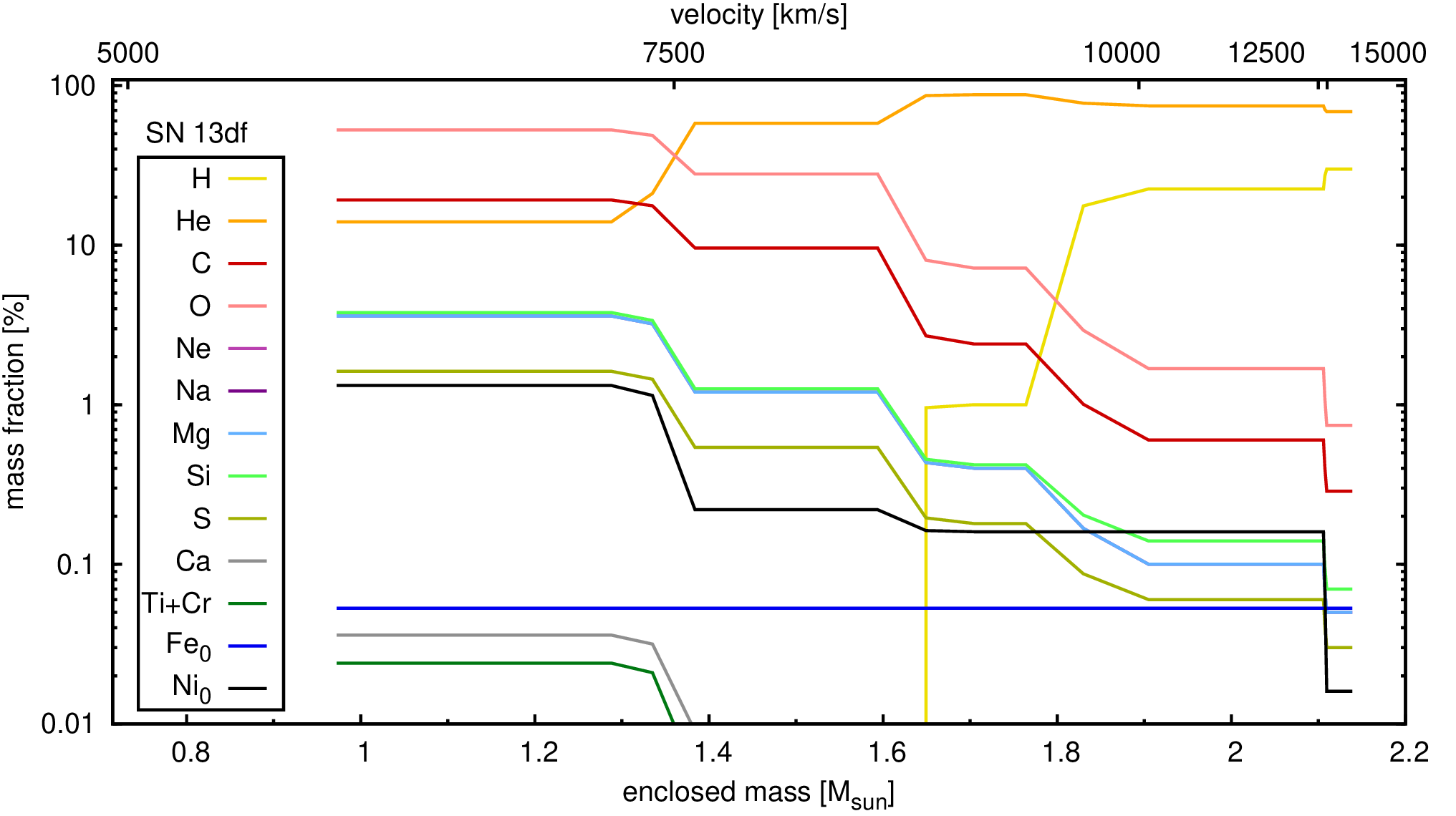}} 
\caption{\small SN~2013df mass stratification for the 30 June 2013 model. Only the outermost 0.1--0.2\,\Msun\ of our envelope is H-enriched; the material below consists mostly of He and C/O in the deeper layers. Traces of iron-group elements are also present at high velocities (i.e., above 10,000\,km\,s$^{-1}$).}
\label{fig:Models1}
\end{figure}

The SN~1993J spectrum taken on 15 April 1993 is dominated by H Balmer lines at 11,000 $\pm$ 1000\,km\,s$^{-1}$, He~I $\lambda\lambda$4471, 4921 at 9000 $\pm$1000\,km\,s$^{-1}$, and the Ca~II H\&K lines at 9000\,km\,s$^{-1}$. Iron-group element lines (e.g., Fe~II, Ti~II, and Cr~II) are also present all across the observed spectrum. 
SN~1993J exhibits a smooth continuum below $\sim 3000$\,\AA\ similar to the one observed in SN~2013df, with some features of iron-group elements identified above the continuum; for further details, see \S3.3 as well as Baron et al. (1993).

The SN~2011dh spectrum taken on 24 June 2011 is dominated by Ca~II H\&K and the Ca~II near-IR triplet at $9000 \pm 400$\,km\,s$^{-1}$, H Balmer lines at velocities of 10,000 $\pm$ 250\,km\,s$^{-1}$, He~I lines (3888, 4471, 4921, 5015, 5875, 6678, and 7065\,\AA) at velocities of  $7600\pm200$\,km\,s$^{-1}$, and the O~I triplet (7774\,\AA) at a velocity of 6700\,km\,s$^{-1}$. 
Contributions from Fe, Ti, and other iron-group elements are also evident all across the observed spectrum at velocities of $7500\pm500$\,km\,s$^{-1}$. At UV wavelengths, the SED is smooth and featureless as in SN~2013df and SN~1993J, but the flux excess is significantly lower than in the latter two cases (Fig. 1, bottom-left panel). 

SN~2001ig spectra taken on 14 and 22 December 2001 (11 and 19\,days past the estimated explosion date) are dominated by He~I lines (3888, 4471, 4921, 5012, 5875\,\AA), Fe~II $\lambda$5169, and H Balmer lines. The contribution from Ca~II to the feature around 3900\,\AA\ can also be inferred from the analysis done on the optical and near-IR spectra (Silverman et al. 2009). At shorter wavelengths, absorption by iron-group elements, mainly Fe~II and Co~II lines, dominates the SED.
All of the observed features strengthen in the later spectrum. Heavy-element lines point to a high expansion velocity of the ejecta ($\sim$\,10,000\,km\,s$^{-1}$), while the H and He lines are observed at velocities of $\sim$\,13,000\,km\,s$^{-1}$. 

\subsection{Spectral Models}
In order to study the ejecta structure of SN~2013df (abundances, density profile), we calculate radiative-transfer models. In addition to modeling the four UV-optical spectra we have of SN~2013df, we also modelled the other SNe in our UV spectroscopic sample.

Modeling SN~2013df, we assume approximate spherical symmetry and follow the ``abundance tomography'' approach of Mazzali et al. (2014). We assume a density profile for the SN and then infer an abundance stratification from photospheric spectra in the phase of homologous, force-free expansion. Doing this, we exploit the fact that the photosphere recedes with time. Early-time spectra show the imprint of the outermost layers of the envelope, and later spectra are dominated by material further inward. Thus, starting from reasonable initial values, the abundances can be inferred step by step. The earliest spectrum is used to determine optimum-fit abundances in the outer layers, and later spectra to subsequently find the abundances in the layers below, while keeping the abundances in the outer layers fixed. This process requires iteration in order to be optimized. We have repeated the fitting procedure several times with different density profiles in order to ensure that the profile finally used reasonably describes the actual envelope.

The spectral synthesis is performed with our spherically symmetric Monte Carlo radiative-transfer code \cite{Mazzali1993,Lucy1999,Mazzali2000}, including abundance stratification \cite{Stehle2005} and a module that calculates the ionization of H and He in full non-local thermodynamic equilibrium (NLTE; Hachinger et al. 2012). The code calculates the radiation field within a spherically symmetric SN expanding ``atmosphere,'' producing synthetic spectra. As input parameters, it reads the abundance and density stratification, as well as the luminosity of the SN,  and the radius/velocity of the ``photosphere'' (the lower boundary of the atmosphere) at each epoch $t$. The code simulates the propagation of energy packets, starting from the photosphere where a blackbody spectrum is assumed to be emitted. The packets interact with atomic lines and free electrons. The state of the plasma, which is needed to calculate these interactions, is computed in NLTE for H and He. Our NLTE module \cite{Hachinger2012} takes into account heating by Compton electrons produced by $\gamma$-rays from \Nifs\ and \Cofs\ decay, using the methods of Lucy (1991). The required heating rates are computed using a modified version of the light-curve code developed by P.\,A.\,Mazzali and used by Cappellaro et al. (1997). In order to arrive at a consistent solution for both the radiation field and plasma, the radiation Monte Carlo routine and the plasma state routine of the code are iteratively called in turn. After convergence, an output spectrum is calculated from the formal integral of the transfer equation.

The density model we use for SN\,2013df is derived from model 13C of Woosley et al. (1994), who set up a model sequence to explain the prototypical SN~1993J. Model 13C has been successfully used by Tsvetkov et al. (2009) to model SN~2008ax. It turns out that the high-velocity tail of 13C produces somewhat too strong H lines. Therefore, we tried different ad-hoc modifications of the model (mostly downscaling the density above a certain threshold velocity/radius). A model that describes SN~2013df well was obtained by scaling the density above 11,000\,\kms\ to one half of its original value; see Figure 4. The original 13C model \cite{Woosley1994} assumes a progenitor mass of 3.7\,\Msun\ (1.5\,\Msun\ of which end up in the remaining neutron star), and our model is only slightly less massive (by 0.04\,\Msun). The explosion energy in 13C was 1.2\,$\times$\,10$^{51}$\,erg, which is reduced to 1.1\,$\times$\,10$^{51}$\,erg in our model. Our estimate of the progenitor mass at the time of explosion is higher than the one derived using [O~I] $\lambda\lambda$6300, 6364 luminosities by Morales-Garoffolo et al. (2014).
We argue that the overall characteristics of SN~2013df are more in agreement with SN~1993J than with SN~2011dh (e.g., SED and duration of shock-cooling phase), and therefore a progenitor mass similar to that of SN~1993J is more plausible. Moreover, our explosion-energy estimate is similar to that derived by Morales-Garoffolo et al. (2014), and assuming a lower progenitor mass at the time of explosion results in higher expansion velocities in our models than the expansion velocities observed.

\begin{figure}[h!p!]
\scalebox{0.5}{\includegraphics[angle=270]{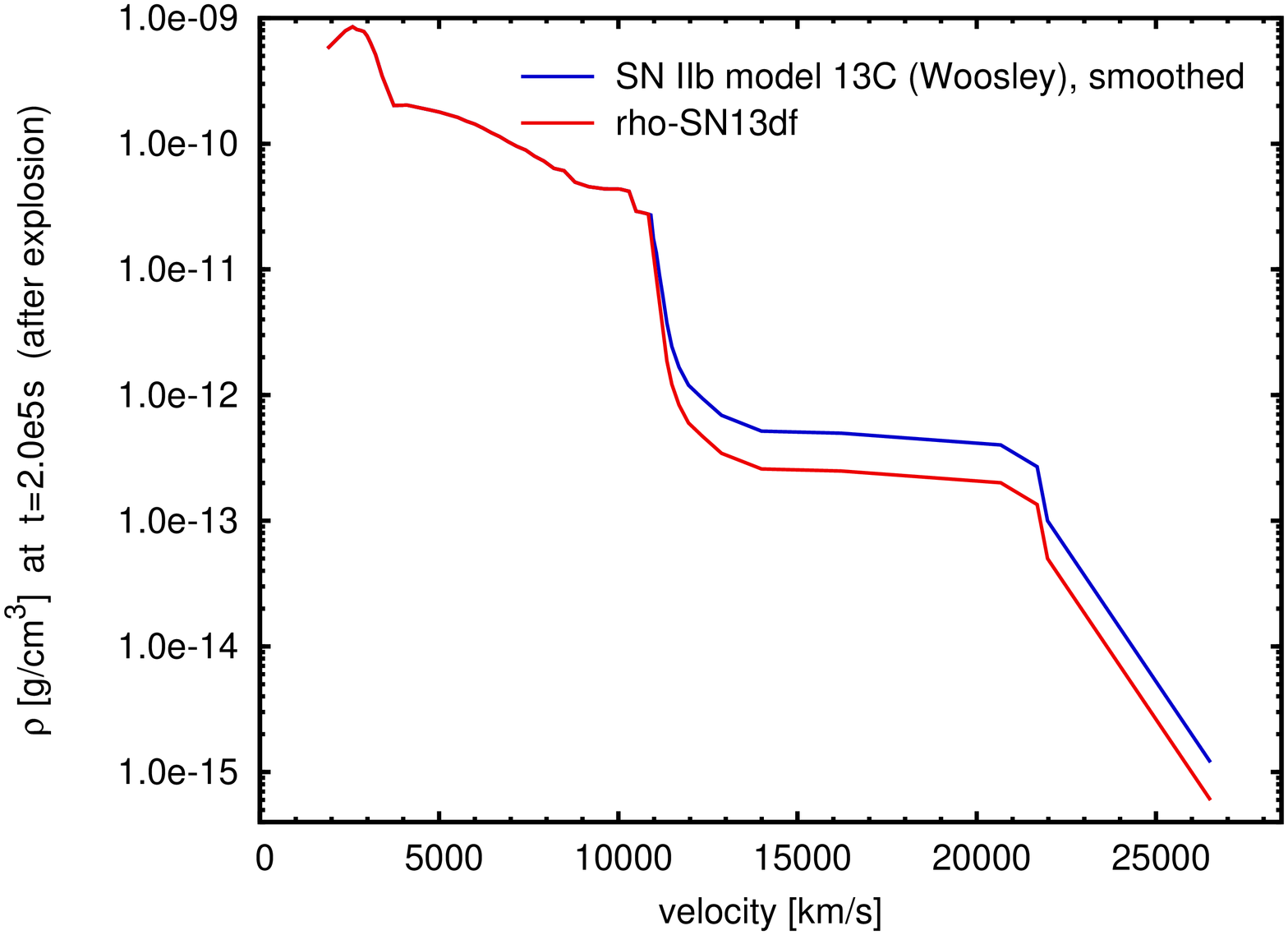}} 
\caption{\small An SN~2013df model density profile at $2\times10^5$\,s ($\sim2.3$\,d) after explosion. A model that well describes SN~2013df was obtained by scaling the density above 11,000\,\kms\ to one half of its original value (red curve).}
\label{fig:Models1}
\end{figure}

The abundances we assume as a starting point are taken from adequate zones of the models for SN~2008ax presented by Hachinger et al. (2012). Only the outermost 0.1--0.2\,\Msun\ of our envelope are H-enriched; the material below mostly consists of He (and C/O in the deeper layers). Taking into account the lower density profile we adopt (Fig. 4), this suggests a lower H-envelope mass for SN~2013df in comparison to SN~1993J. Under these constraints, we optimized the abundances, inferring the best-fitting values. The synthetic spectra obtained using our radiative-transfer code are given in Figure 5. 
Our models suggest a blackbody temperature of 8200\,K and an effective photosphere temperature of 5650\,K in the spectrum taken on 17 June 2013, dropping to a blackbody temperature of 6800\,K and an effective photosphere temperature of 5300\,K in the spectrum taken on 30 June 2014, in agreement with Morales-Garoffolo et al. (2014). The photospheric velocity also drops from 9600\,km\,s$^{-1}$ to 5900\,km\,s$^{-1}$.
Velocity-space stratification for the model describing the spectrum taken on 30 June 2013 is given in Figure 3. Only the outermost 0.1--0.2\,\Msun\ of our envelope are H-enriched; the material below mostly consists of He and C/O in the deeper layers. Traces of iron-group elements are also present at high velocities (i.e., above 10,000\,km\,s$^{-1}$).

We find a good match between the observed and synthetic spectra at wavelengths above $\sim5000$\,\AA. At shorter wavelengths, our models simulate most of the observed lines (mainly Fe~II and Ca~II), but the observed spectrum seems to include an additional component that is not affected by line absorption and is responsible to most of the emitted flux. This trend is consistent in all 4 simulated spectra, and is stronger at shorter wavelengths. In an attempt to increase the UV flux, which in our models is absorbed by iron-group elements in the UV (Fe, Ti, Cr), we have lowered the metal content in the outer layers of the ejecta, but were unable to reproduce the observed iron-group element lines above $\sim4000$\,\AA.  Increasing the metal content in the outer layers to allow higher probability for reverse fluorescence induces line structure in the UV that is not present in the observed spectra. We therefore conclude that the source of the UV emission is above the photosphere, and is most likely associated with CSM shocked by the SN blast wave. This argument is supported by the flat UV LC after the first peak, as well as by the nebular spectral analysis, both presented by Morales-Garoffolo et al. (2014) .

\begin{figure}[h!p!]
\centerline{
\begin{tabular}{cc}
\scalebox{0.32}{\includegraphics{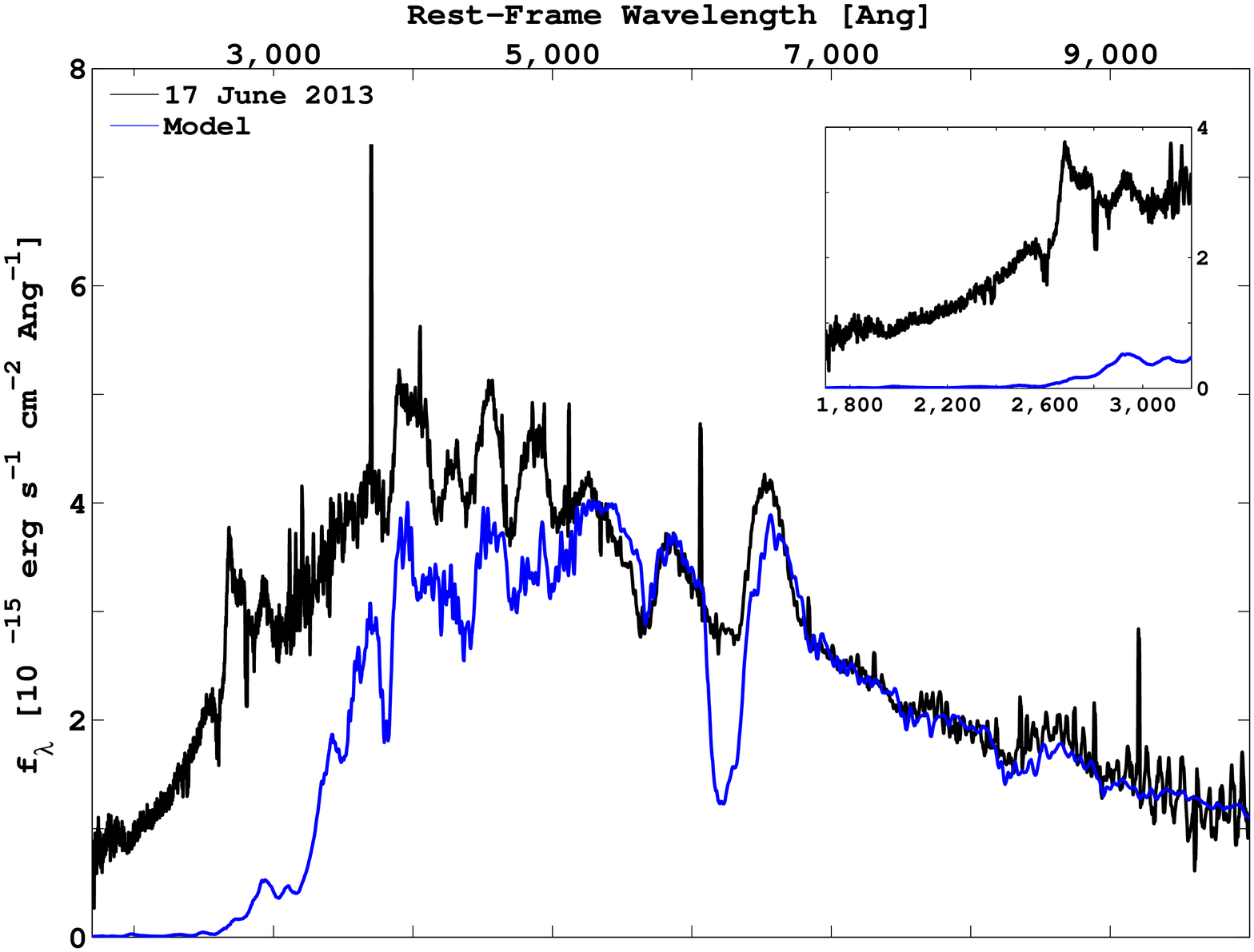}} &
\scalebox{0.32}{\includegraphics{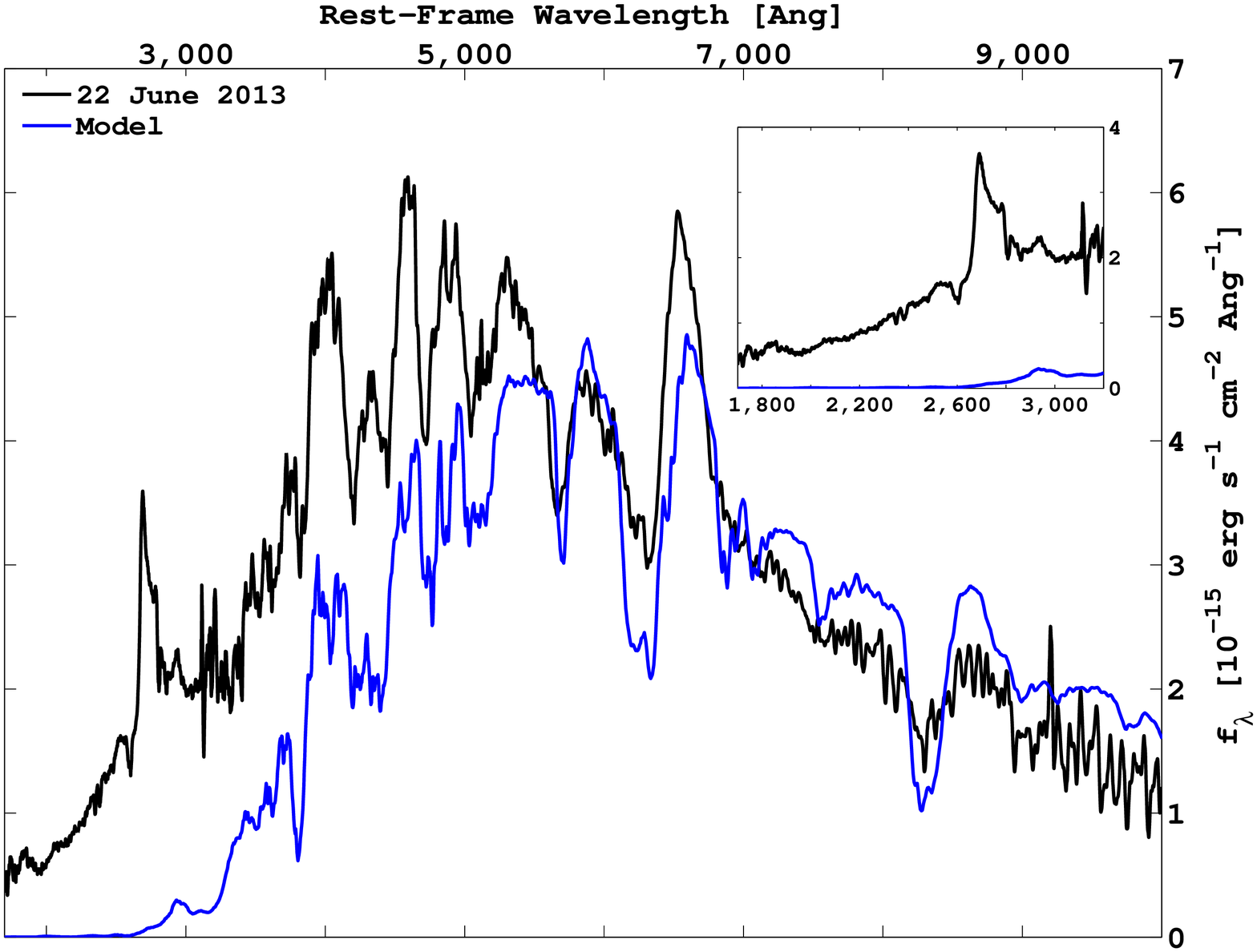}} \\
\scalebox{0.32}{\includegraphics{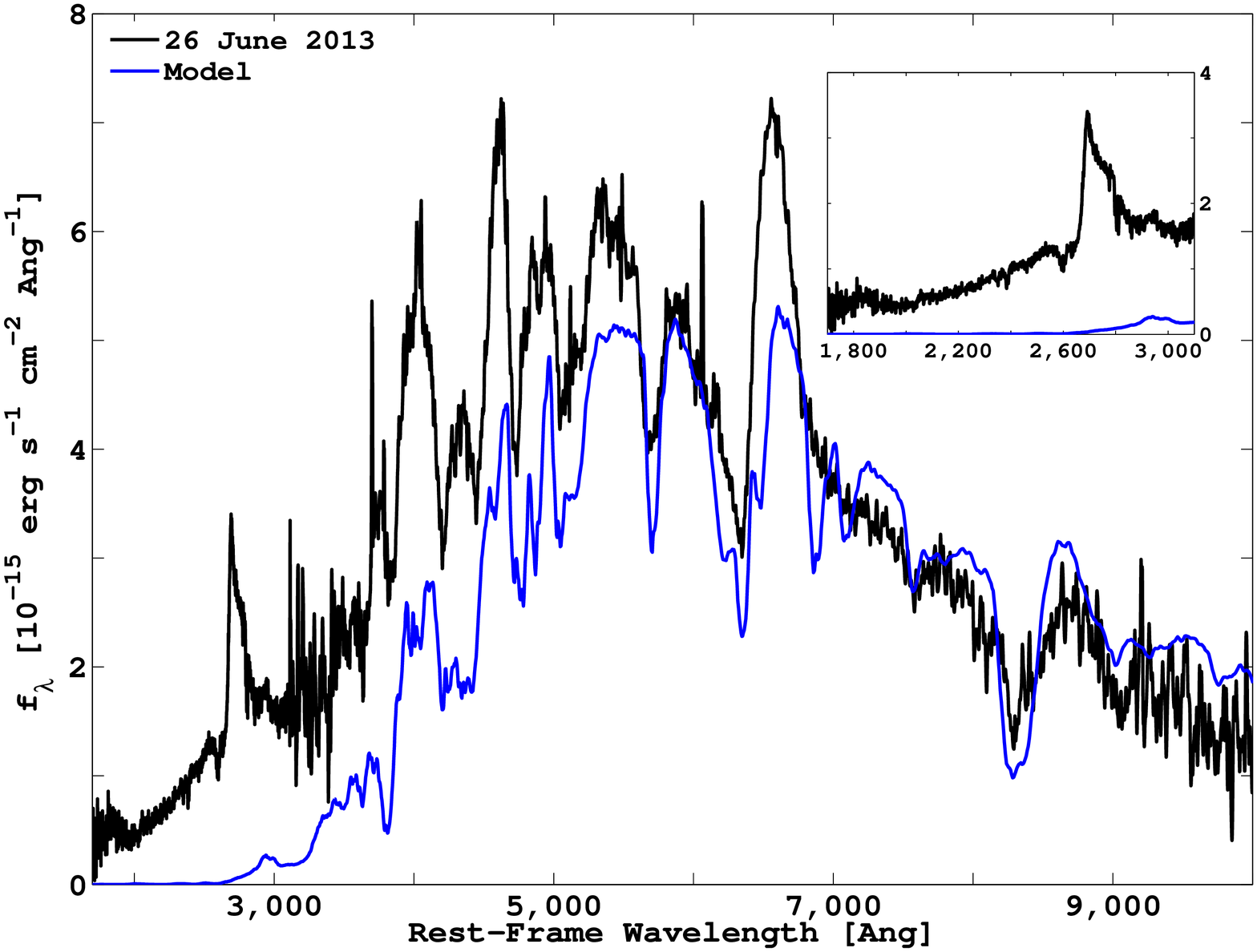}} &
\scalebox{0.32}{\includegraphics{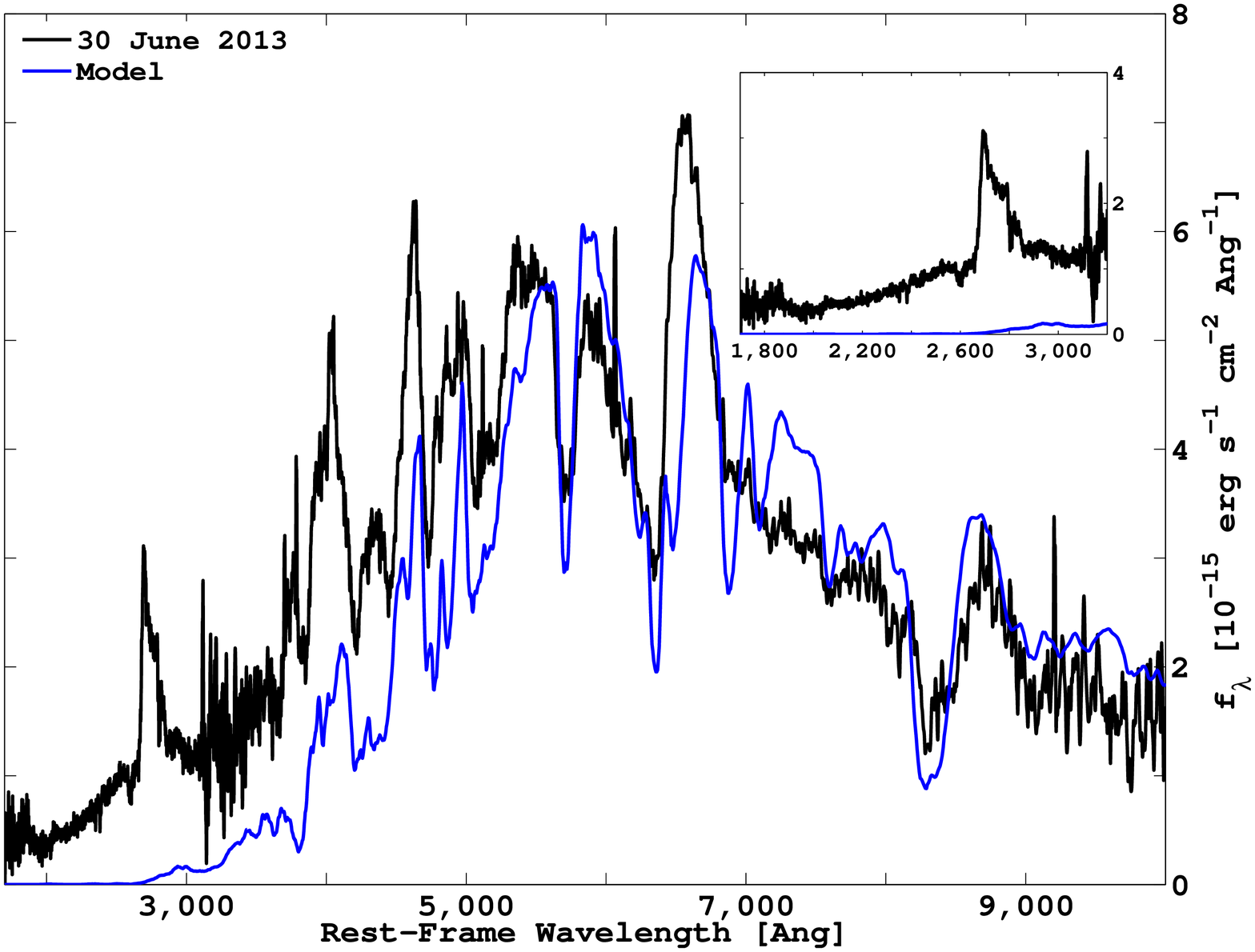}} \\
\end{tabular}}
\caption{\small SN~2013df synthetic spectra. We get a good match above 5000\,\AA. At shorter wavelengths, an additional component that is not affected by line absorption from iron-group elements seems to be present. We associate this component with emission from shocked CSM above the photosphere.}
\label{fig:Models1}
\end{figure}

The models for SNe 1993J, 2001ig, and 2011dh are set up as minimal modifications of our SN\,2013df model. We have chosen this strategy, partly justified by the observed similarities in the bolometric LCs (\S3.1), so as to bring out the principal physical changes from one to the other supernova -- even if more detailed differences in the density structure, etc., between the SNe may thus not be captured in our study.

With this approach, one spectrum of each comparison SN is sufficient in order to allow for a basic comparison. We started out rerunning the SN~2013df model with the modified values for epoch and basic SN parameters (such as distance and reddening), and a starting guess for photospheric velocity and luminosity. Usually, we arrived at a well-fitting model upon further optimizing the latter two parameters, and only slightly changing the abundance stratification.

For SN~1993J, a detailed analysis of the spectrum based on the Woosley et al. (1994) 13C model we have used can be found in Woosley et al. (1994). The model suggests a blackbody temperature of 8200\,K, a photosphere effective temperature of 5900\,K, and a photosphere velocity of 8250\,km\,s$^{-1}$.
For SN~2011dh, our models suggest a blackbody temperature of 7470\,K and an effective photospheric temperature of 5900\,K, slightly lower than the results of Ergon et al. (2014a). The photospheric velocity we infer is 6750\,km\,s$^{-1}$, in agreement with Ergon et al. (2014a).

For both SN~1993J and SN~2011dh, we find a similar trend to the one observed for SN~2013df, in which an additional component not affected by line absorption is responsible for most of the emission below $\sim4000$\,\AA. The relative amount of flux attributed to this component is weakest in the case of SN~2011dh, and similar in the case of SN~1993J, as seen in the insets of Figure 6. 

\begin{figure}[h!p!]
\centerline{
\begin{tabular}{cc}
\scalebox{0.3}{\includegraphics{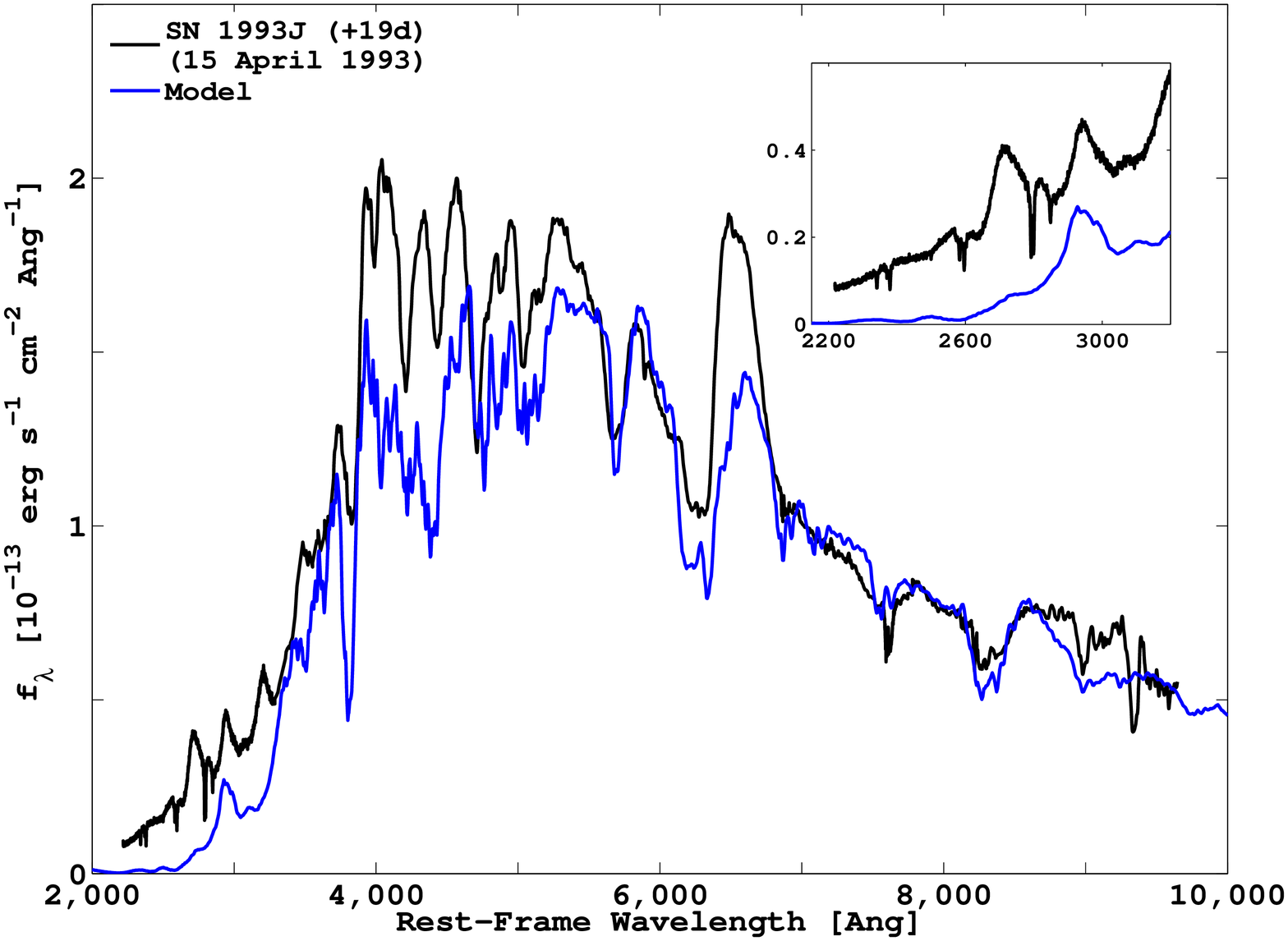}} &
\scalebox{0.3}{\includegraphics{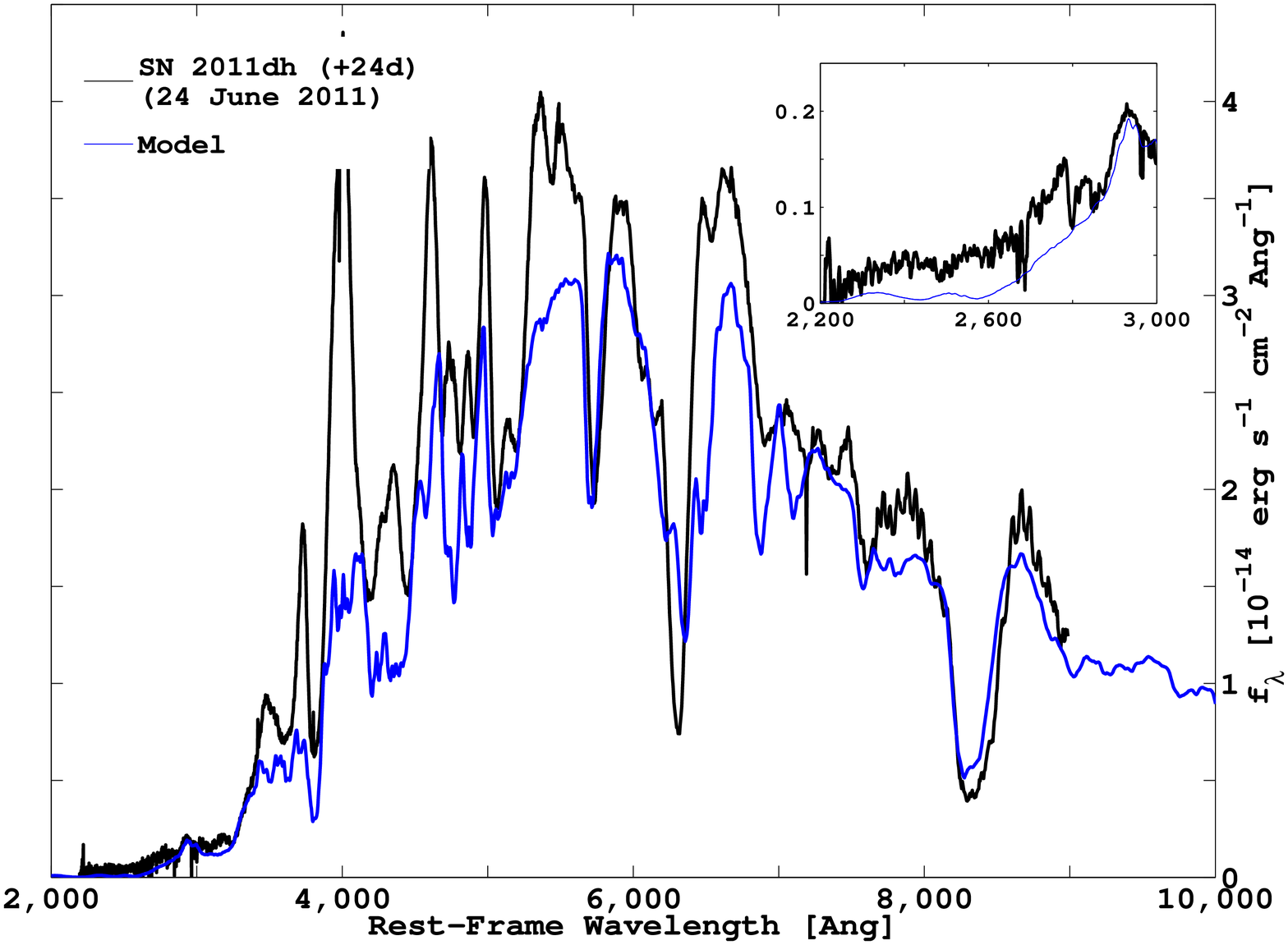}} \\
\end{tabular}}
\caption{\small  SN~1993J and SN~2011dh synthetic spectra. We obtain similar results to those for SN~2013df. The amount of flux excess compared to our synthetic models varies between events.}
\label{fig:Models2}
\end{figure}

\begin{figure}[h!p!]
\center
    \scalebox{0.35}{\includegraphics{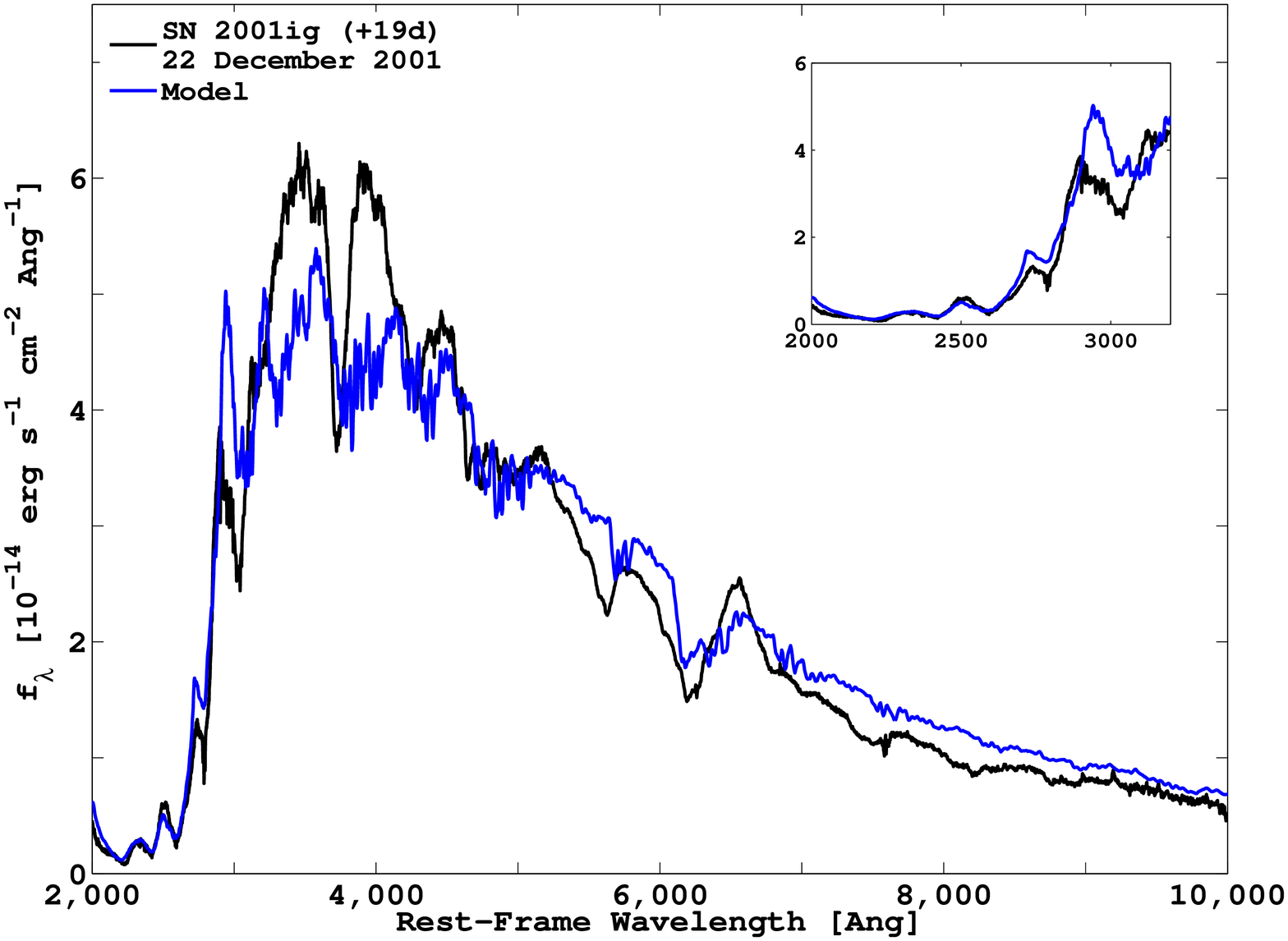}}
  \caption{\small A SN~2001ig synthetic spectrum. We achieve a good match at all wavelengths. The mechanism producing the observed UV flux is reverse fluorescence of photons at longer wavelengths by excited iron-group element atoms. The same mechanism is responsible for the observed UV emission in SNe~Ia (e.g., Hachinger et al. 2012).
          }
\end{figure}

In the case of SN~2001ig, our models suggest a photosphere velocity of 8750\,km\,s$^{-1}$, a photospheric temperature of 7000\,K, and a blackbody temperature of 12,000\,K, in agreement with Silverman et al. (2009). The UV flux ratio below 3000\,\AA\ is much lower. Iron-group element lines are clearly seen in this wavelength range, and we obtain a good match between our synthetic spectrum and the observed one at all wavelengths; see Figure 7.
In addition, the line strength in the UV is directly proportional to the flux at longer wavelengths, as seen by comparing the spectra taken on 14 December 2001 (11 days after estimated explosion) and on 22 December 2001 (19 days after estimated explosion); see the bottom-right panel of Figure 1. 
We conclude that the SN~2001ig UV flux originates from reverse fluorescence of photons at longer wavelengths by excited iron-group elements in the expanding ejecta, and is similar to the observed UV spectrum of Type Ia SNe (e.g., Panagia 2007; Hachinger et al. 2012; Mazzali et al. 2014); see also Silverman et al. (2013) for a discussion of Type Ia SNe interacting with CSM. As mentioned in \S3.1, this SN is intrinsically the brightest event in our sample, and it synthesized the most $^{56}$Ni \cite{Silverman2009}.

\newpage
\section{Discussion}
Prior work has investigated diversity in the optical spectra and in emission at X-ray and radio bands among Type IIb SNe.  Based on these observables, Chevalier \& Soderberg (2010) claim there is an internal division between SNe~IIb having compact progenitors and those with extended progenitors. Studies of pre-explosion images from the {\it HST} archive show that indeed SN~1993J, SN~2011dh, and SN~2013df had YSG progenitors (e.g., Aldering et al. 1994; Maund et al. 2011; Van Dyk et al. 2013, 2014; Ergon et al. 2014a.), while observations of the SN~2008ax progenitor are consistent with a compact Wolf-Rayet star, though this is inconclusive \cite{Crockett2008}. Ryder et al. (2004) suggest that SN~2001ig might also belong to the class of SNe~IIb with a compact progenitor, based on analysis of data in the radio waveband. 

In this study we have examined early-time UV spectra of Type IIb SNe in an attempt to find whether the UV band holds information that might further reveal the nature of SN~IIb progenitors and their vicinity. We have gathered a sample of four events for which high-quality UV spectra were obtained within 30\,days of the time of estimated explosion, and using radiative-transfer models we investigate the spectral diversity in the UV band between these events. 

We find that our sample can be divided into two groups. The first group includes SN~1993J, SN~2011dh, and SN~2013df, all associated with extended YSG progenitors ($\sim 10^{13}$\,cm), with expected low-velocity winds (e.g., Bufano et al. 2014, and reference therein). For these three events we find a UV continuum flux excess, not affected by line absorption. We conclude it most likely originates above the photosphere and is the result of shocked low-metallicity CSM. The flux excess varies between events. It is strongest in the case of SN~2013df and weakest in the case of SN~2011dh. 
Comparing the duration of the shock-cooling phase among the three SNe, we find a correlation between the shock-cooling phase duration and the amount of flux in the UV attributed to the shocked CSM:  the duration of the shock-cooling phase is longest in the case of SN~2013df, and shortest for SN~2011dh. This seems to suggest that the shock-cooling phase is affected not only by the relative amount of mass at the outer layers of the progenitor \cite{Nakar2014}, but also by the presence of dense CSM at the progenitor vicinity at the time of explosion.
The presence of dense CSM will also affect the blast-wave velocity derived through observed peak spectral radio luminosities\footnote{A unique event in this context is SN~2003bg \cite{Hamuy2009,Mazzali2009}, which exhibits a high blast-wave velocity attributed to the presence of CSM, but with no indication for an initial shock-cooling phase. SN 2003bg was classified as a Type IIb broad-line SN, and is more energetic than the events analyzed in this work --- further expanding the diversity of Type IIb SNe.} \cite{Chevalier2010,Horesh2013,Bufano2014}.
A comparison reveals that SN~2011dh had a higher velocity than SN~1993J, while SN~2013df had a velocity similar\footnote{Private correspondence with A. Soderberg and A. Kamble. Radio data of SN~2013df are not public at the time of the writing of this paper; see a future paper by Kamble et al.} to the one measured for SN~1993J. A comparison between SN~2011dh and SN~2011hs further illustrates the diversity in this class of events. We suggest that SN~2011hs is an intermediate case, with  a CSM amount in between that of SN~2011dh, and SN~2013dh. This interpretation is supported by the cooling-phase duration and the inferred blast-wave velocity (Table 1), but is speculative because the lack of early-time UV spectra of SN~2011hs does not allow us to reach a clear conclusion. Future observations of Type IIb SNe (specifically, early-time UV spectroscopy) will allow us to develop a more quantitative understanding of the effect of CSM on observables such as shock cooling phase duration and blast-wave velocity.

In the case of SN~2001ig, the observed UV spectra display a weak continuum and strong reverse-fluorescence features, similar to the UV spectra of Type Ia SNe such as SN~1992A \cite{Kirshner1993}. A comparison of $V$-band LCs shows that indeed SN~2001ig was the most luminous event in our sample, and so the photon budget in the visible band was higher, leading to a higher probability for reverse fluorescence to occur. We argue that the observed similarity to SNe~Ia reflects a high ratio of radioactive nickel mass to total mass and supports the inclusion of this event among the group of SNe~IIb originating from a compact progenitor. 

Our analysis suggests a correlation between the UV SED, blast-wave velocity, shock-cooling phase, and ejected $^{56}$Ni mass; see Table 1. The first three observables seem to be affected by the amount of CSM around the progenitor at the time of explosion, while the amount of $^{56}$Ni mass ejected affects the shape of the SED (i.e., quasi-continuum vs. line-dominated). The observed correlation might lead to wrong estimates of progenitor radii. In the case of SN~2013df, our hypothesis explains the slight discrepancy in progenitor radius estimates observed by Van Dyk et al. (2014), as one method used is affected by the presence of CSM (i.e., progenitor radius based on shock-cooling phase), while the other is not (i.e., progenitor radius based on {\it HST} images). In a similar manner, the presence of circumstellar material can explain the large discrepancy between the progenitor radius estimates detailed by Morales-Garoffolo et al. (2014).

Our findings extend those of Chevalier \& Soderberg (2010), showing the diversity within the group of SNe~IIb having an extended progenitor, and add emission in the UV band to the list of observables that might help us in understanding this diversity. 
Most notably, we show a correlation between the UV SED and observables related to the progenitor and its environment. Our findings are limited by our small sample size, in which only a single event is assumed to belong to the class of SNe~IIb originating from a compact progenitor. Future work should therefore aim to increase the number of SNe~IIb for which we do have early-time high-quality UV data simultaneously with data across the electromagnetic spectrum in an attempt to better constrain the type of progenitors and the processes that lead to this important class of CC~SNe.   

\newpage
\section*{Acknowledgments}
\noindent S.B acknowledges support for this work by NASA through Einstein
Postdoctoral Fellowship grant number PF0000 awarded by the
Chandra X-ray Center, which is operated by the Smithsonian
Astrophysical Observatory for NASA under contract NAS8-03060.
S.H is supported by a Minerva ARCHES award.
A.G. acknowledges support by grants from the ISF, BSF, GIF, Minerva, FP7/ERC
grant \#307260, the ``Quantum-Universe" I-core program of the planning and budgeting
committee and the ISF, and a Kimmel Investigator award.
A.V.F. and his group at UC Berkeley have depended on generous
financial assistance from Gary \& Cynthia Bengier, the Richard \&
Rhoda Goldman Fund, the Sylvia \& Jim Katzman Foundation, the
Christopher R. Redlich Fund, the TABASGO Foundation, NSF grant
AST-1211916, and NASA/HST grant GO-13030 from the Space Telescope
Science Institute (which is operated by the Association of
Universities for Research in Astronomy, Inc., under NASA contract NAS
05-26555).
J.M.S is supported by an NSF Astronomy and Astrophysics Postdoctoral Fellowship under award AST-1302771.

\end{document}